# MVPBT: R package for publication bias tests in meta-analysis of diagnostic accuracy studies

Hisashi Noma, PhD*
Department of Data Science, The Institute of Statistical Mathematics, Tokyo, Japan
ORCID: http://orcid.org/0000-0002-2520-9949

*Corresponding author: Hisashi Noma
 Department of Data Science, The Institute of Statistical Mathematics
 10-3 Midori-cho, Tachikawa, Tokyo 190-8562, Japan
 TEL: +81-50-5533-8440
 e-mail: noma@ism.ac.jp

**Abstract**

Meta-analysis for diagnostic test accuracy (DTA) has been a standard research method for synthesizing evidence from diagnostic studies. In DTA meta-analysis, although publication bias is an important source of bias, no certain methods similar to the Egger test in univariate meta-analysis have been developed to detect such bias. However, several recent studies have discussed these methods in the framework of multivariate meta-analysis, and some generalized Egger tests have been developed. The R package `MVPBT` (https://cran.r-project.org/web/packages/MVPBT/) was developed to implement the generalized Egger tests developed by Noma (2020; *Biometrics* **76**, 1255-1259) for DTA meta-analysis. Noma's publication bias tests effectively incorporate the correlation information between multiple outcomes and are expected to improve the statistical powers. The present paper provides a nontechnical introduction and practical examples of data analyses of the publication bias tests of DTA meta-analysis using the `MVPBT` package.

## 1. Introduction

Meta-analysis for diagnostic test accuracy (DTA) has been a standard research method for synthesizing evidence from diagnostic studies (Deeks, 2001; Leeflang et al., 2008). In these studies, bivariate meta-analysis methods are widely used to adequately incorporate correlation information between outcome variables (Reitsma et al., 2005). Summarized information of diagnostic accuracy can also be expressed by the summary receiver operating characteristic (SROC) curve (Harbord et al., 2007; Rutter and Gatsonis, 2001).

Publication bias is an important source of bias in these evidence synthesis studies (Higgins and Thomas, 2019). In conventional univariate meta-analysis methodology, although various statistical tests, including the Egger test (Egger et al., 1997), have been developed to detect publication bias, no certain methods for DTA meta-analysis have been developed. However, several studies have recently discussed these methods, and generalized Egger tests have been developed for multivariate meta-analysis (Hong et al., 2020; Noma, 2020). The R package `MVPBT` (https://cran.r-project.org/web/packages/MVPBT/) was developed to implement the generalized Egger tests developed by Noma (2020) for DTA meta-analysis. Noma's publication bias tests effectively incorporate the correlation information between multiple outcomes and are expected to improve the statistical powers. The present work provides a nontechnical introduction and practical examples of data analyses of the publication bias tests of DTA meta-analysis using `MVPBT`.

## 2. The generalized Egger tests

The Egger test (Egger et al., 1997) is a statistical test to detect asymmetry of a funnel plot. A generalized version of this test for multivariate meta-analysis was first developed by Hong et al. (2020) as an overall test for the global null hypothesis "all of the univariate funnel plots for multiple outcomes are symmetric." This overall test incorporates multiple outcome information, and the statistical power is generally improved compared with that of the conventional univariate publication bias test (Hong et al., 2020). Hong's test (known as the multivariate small study effect test [MSSET]) circumvents the incorporation of the correlation information among the multiple outcomes that are sometimes unavailable under certain situations of multivariate meta-analysis. However, for DTA meta-analysis, all of the correlation information is available in the Reitsma's bivariate meta-analysis model (Reitsma et al., 2005); in addition, the statistical power of MSSET might be inefficient because it does not use the correlation information. Noma (2020) developed alternative, efficient, generalized Egger tests for the same global null hypothesis; these tests (referred to as MSSET2 and MSSET3) effectively incorporate the correlation information. Because of their use of correlation information, Noma's tests are expected to improve the statistical powers compared with those of MSSET when applied to DTA meta-analysis. In Table 1, we provide brief nontechnical explanations of these tests.

For the generalized Egger tests, we consider a multivariate meta-analysis with $m$ studies, where $J$ outcomes are of interest (under this specific situation, $J = 2$). We denote the summary measure for the $j$th outcome of $i$th study as $Y_{ij}$ ($i = 1,\ldots,m; j = 1,\ldots,J$), and consider the multivariate random-effects model for the multivariate outcome $\boldsymbol{Y}_i = (Y_{i1}, \ldots, Y_{iJ})^T$,

$$\boldsymbol{Y}_i \sim \text{MVN}(\boldsymbol{\theta}_i, \boldsymbol{\Delta}_i)$$

$$\boldsymbol{\theta}_i \sim \text{MVN}(\boldsymbol{\beta}, \boldsymbol{\Omega}),$$

where $\boldsymbol{\beta} = (\beta_1, \ldots, \beta_J)^T$ is the grand mean vector and $\boldsymbol{\Delta}_i$ and $\boldsymbol{\Omega}$ are $J \times J$ within-study and between-study covariance matrices, respectively. We denote the marginal covariance matrix of $\boldsymbol{Y}_i$ as $\boldsymbol{V}_i = \boldsymbol{\Delta}_i + \boldsymbol{\Omega}$. We also denote the within-study and between-study variances as $s_{i1}^2, \dots, s_{iJ}^2$ and $\tau_1^2, \dots, \tau_J^2$, respectively (the diagnostic elements of $\boldsymbol{\Delta}_i$ and $\boldsymbol{\Omega}$).

For the regression-based publication bias tests, we consider standardized effect sizes, $\text{SND}_{ij} = Y_{ij}(s_{ij}^2 + \tau_j^2)^{-1/2}$ as an outcome quantity, and the joint distribution of $\text{SND}_{ij}$ is

$$\textbf{SND}_i \sim \text{MVN}(\boldsymbol{P}_i \boldsymbol{b}, \boldsymbol{\Psi}_i),$$

where $\textbf{SND}_i = (\text{SND}_{i1}, \dots, \text{SND}_{iJ})^T$, $\boldsymbol{P}_i = \text{diag}(P_{i1}, \dots, P_{iJ})$; $P_{ij} = (s_{ij}^2 + \tau_j^2)^{-1/2}$, and $\boldsymbol{\Psi}_i = \boldsymbol{P}_i \boldsymbol{V}_i \boldsymbol{P}_i^T$ ($i = 1, \dots, m$). We then consider the following multivariate regression model:

$$\textbf{SND}_i = \boldsymbol{a} + \boldsymbol{P}_i \boldsymbol{b} + \boldsymbol{\varepsilon}_i, \tag{*}$$

where $\boldsymbol{a} = (a_1, \dots, a_J)^T$, $\boldsymbol{b} = (b_1, \dots, b_J)^T$, and $\boldsymbol{\varepsilon}_i = (\varepsilon_{i1}, \dots, \varepsilon_{iJ})^T$; $\boldsymbol{\varepsilon}_i \sim \text{MVN}(\boldsymbol{0}, \boldsymbol{\Psi}_i)$. The MSSET, MSSET2, and MSSET3 tests were derived as pseudo-score tests for a null hypothesis $H_0$: $\boldsymbol{a} = \boldsymbol{0}$ vs. $H_1$: $\boldsymbol{a} \neq \boldsymbol{0}$ (at least one component of $\boldsymbol{a}$ is nonzero) as a natural extension of the Egger test.

MSSET2 is the score test for the null hypothesis of $H_0$: $\boldsymbol{a} = \boldsymbol{0}$. The score statistic is written as

$$\boldsymbol{T} = \boldsymbol{U}_{\boldsymbol{a}}(\boldsymbol{0}, \tilde{\boldsymbol{b}})^T \boldsymbol{J}_{\boldsymbol{a}}(\boldsymbol{0}, \tilde{\boldsymbol{b}})^{-1} \boldsymbol{U}_{\boldsymbol{a}}(\boldsymbol{0}, \tilde{\boldsymbol{b}}),$$

where

$$\boldsymbol{U}_{\boldsymbol{a}}(\boldsymbol{a}, \boldsymbol{b}) = \sum_{i=1}^m \boldsymbol{\Psi}_i^{-1}(\textbf{SND}_i - \boldsymbol{a} - \boldsymbol{P}_i \boldsymbol{b}),$$

$$\boldsymbol{J}_{\boldsymbol{a}}(\boldsymbol{a}, \boldsymbol{b}) = \left(\sum_{i=1}^m \boldsymbol{\Psi}_i^{-1}\right) - \left(\sum_{i=1}^m \boldsymbol{P}_i^T \boldsymbol{\Psi}_i^{-1}\right) \left(\sum_{i=1}^m \boldsymbol{P}_i^T \boldsymbol{\Psi}_i^{-1} \boldsymbol{P}_i\right)^{-1} \left(\sum_{i=1}^m \boldsymbol{\Psi}_i^{-1} \boldsymbol{P}_i\right),$$

and $\tilde{\boldsymbol{b}}$ is the constrained maximum likelihood estimator (CMLE) of $\boldsymbol{b}$ under $H_0$,

$$\tilde{\boldsymbol{b}} = \left(\sum_{i=1}^m \boldsymbol{P}_i^T \boldsymbol{\Psi}_i^{-1} \boldsymbol{P}_i\right)^{-1} \left(\sum_{i=1}^m \boldsymbol{P}_i^T \boldsymbol{\Psi}_i^{-1} \textbf{SND}_i\right).$$

The score statistic **T** approximately follows the $\chi^2$ distribution with $J$ degrees of freedom. Also, the MSSET3 test is constructed as the bootstrap testing method for the score statistic.

**Algorithm: MSSET3**

1. For the multivariate regression model (*), compute the CMLE of **b** under H$_0$, $\tilde{\mathbf{b}}$.

2. Resample $\mathbf{SND}_1^{(k)}, \mathbf{SND}_2^{(k)}, \ldots, \mathbf{SND}_m^{(k)}$ from the estimated null distribution of (*), MVN($\mathbf{P}_i\tilde{\mathbf{b}}, \mathbf{\Psi}_i$), via parametric bootstrap, $B$ times ($k = 1,2,\ldots,B$).

3. Compute the bootstrap estimate of $\mathbf{\Omega}$ for the multivariate meta-analysis model (3), $\tilde{\mathbf{\Omega}}^{(k)}$ by the $k$th bootstrap sample $\mathbf{SND}_1^{(k)}, \mathbf{SND}_2^{(k)}, \ldots, \mathbf{SND}_m^{(k)}$. Then compute the score statistic $T_2^{(k)}$ by the $k$th bootstrap sample $\mathbf{SND}_1^{(k)}, \mathbf{SND}_2^{(k)}, \ldots, \mathbf{SND}_m^{(k)}$, replacing $\mathbf{\Omega}$ in $\mathbf{\Psi}_i$ with $\tilde{\mathbf{\Omega}}^{(k)}$. Also, replicate this procedure for all $B$ bootstrap samples.

4. The bootstrap estimate of the sampling distribution of **T** can be obtained by the empirical distribution of $\mathbf{T}^{(1)}, \mathbf{T}^{(2)}, \ldots, \mathbf{T}^{(B)}$. Use the bootstrap distribution as the reference distribution of the score test statistic **T**.

The approximation of the sample distribution of **T** is expected to be improved by using the bootstrap distribution as the reference distribution.

### 3. MVPBT: Generalized Egger tests for DTA meta-analysis

*3.1 Installation and preparation of MVPBT*

MVPBT is available from the CRAN (The Comprehensive R Archive Network) and can be downloaded through the standard installation command of R:

```
> install.packages("MVPBT")
```

The details of the package and its manual is also available at the CRAN page [https://cran.r-project.org/web/packages/MVPBT/].

### 3.2 Illustrative example: DTA meta-analysis for lymphangiography

As an illustrative example, we present the result of the pooling analysis for a DTA meta-analysis that assessed the diagnostic accuracy of lymphangiography based on the presence of nodal-filling defects for the diagnosis of lymph-node metastasis in women with cervical cancer (Scheidler et al., 1997; $N = 17$) in Table 2 and the SROC curve in Figure 1(a). We also provided funnel plots for the two outcome variables in the Reitsma-type bivariate meta-analysis model, i.e., logit-transformed sensitivity (logit(Se)) and the logit-transformed false-positive rate (logit(FPR)) in Figure 1(b) and (c), respectively. The global null hypothesis of Noma's (2020) publication bias tests is "both of the two univariate funnel plots are symmetric." Thus, these tests are interpreted as statistical tests to assess asymmetries of the two funnel plots jointly. The *P*-values of the univariate Egger tests were 0.551 and 0.029, and asymmetry of the funnel plot was indicated for the FPR.

### 3.3 Data preprocessing

Firstly, some data preprocessing is required to perform the publication bias tests. For DTA meta-analysis, the original data of individual studies are typically reported as numbers of true positives (TP), false positives (FP), false negatives (FN), and true negatives (TN). These outcome data should be transformed to summary statistics that are suitable to fit the Reitsma's bivariate meta-analysis model. In the `MVPBT` package, a useful function `edta` is involved, that performs this calculation. A summary of `edta` function is presented in Table 3. Also, an example is as follows:

```
> data(cervical)
> LAG <- cervical[cervical$method==2,]
> attach(LAG)
> edta(TP,FN,TN,FP)
```

$y

|        | Y1         | Y2         |
|--------|------------|------------|
| [1,]   |  0.6190392 | -3.9951379 |
| [2,]   |  1.2237754 | -0.3513979 |
| [3,]   |  1.1999648 | -3.4965076 |
| [4,]   |  0.7884574 | -2.5123056 |
| [5,]   |  0.3364722 | -1.0399444 |
| [6,]   |  1.2237754 | -1.6094379 |
| [7,]   |  1.2992830 |  0.1251631 |
| [8,]   |  0.2984930 | -1.0986123 |
| [9,]   |  0.6632942 | -0.7563261 |
| [10,]  |  0.5877867 | -1.0986123 |
| [11,]  | -0.2113091 | -1.7298841 |
| [12,]  |  0.8472979 | -2.5123056 |
| [13,]  | -0.9555114 | -1.4522523 |
| [14,]  |  0.0000000 | -1.0663514 |
| [15,]  |  1.2697605 | -0.9869983 |
| [16,]  |  1.7346011 | -2.3715780 |
| [17,]  |  2.1972246 | -1.4213857 |

$S

|        | V1         |   | V2         |
|--------|------------|---|------------|
| [1,]   | 0.14652015 | 0 | 0.67893661 |
| [2,]   | 0.51764706 | 0 | 0.17933723 |
| [3,]   | 0.10409639 | 0 | 0.68686869 |
| [4,]   | 0.58181818 | 0 | 0.72072072 |
| [5,]   | 0.05274725 | 0 | 0.02313631 |
| [6,]   | 0.51764706 | 0 | 0.18461538 |
| [7,]   | 0.84848485 | 0 | 0.25098039 |
| [8,]   | 0.15147265 | 0 | 0.07619048 |
| [9,]   | 0.17825312 | 0 | 0.12777285 |
| [10,]  | 0.62222222 | 0 | 0.15686275 |

```
[11,] 0.21288515 0 0.09418440
[12,] 0.31746032 0 0.24024024
[13,] 0.55384615 0 0.22437137
[14,] 0.26666667 0 0.12802498
[15,] 0.10247191 0 0.02718205
[16,] 0.78431373 0 0.31238095
[17,] 2.22222222 0 0.35467980

$Se
 [1] 0.6500000 0.7727273 0.7685185 0.6875000 0.5833333 0.7727273 0.7857143
 0.5740741 0.6600000 0.6428571 0.4473684 0.7000000 0.2777778 0.5000000
 0.7807018 0.8500000 0.9000000

$Fp
 [1] 0.01807229 0.41304348 0.02941176 0.07500000 0.26116071 0.16666667
 0.53125000 0.25000000 0.31944444 0.25000000 0.15060241 0.07500000
 0.18965517 0.25609756 0.27150538 0.08536585 0.19444444
```

*3.4 The generalized Egger tests: MSSET2 and MSSET3*

There are two options for the publication bias tests for the global null hypothesis of asymmetries of two funnel plots in the `MVPBT` package, i.e., the MSSET2 an MSSET3. In the `MVPBT` package, they are implementable by `MVPBT2` and `MVPBT3` functions, respectively. Details of the methods were described in the Section 2.

The former test (MSSET2) can be implemented by the following command. The augments `y` and `S` are computable by `edta` function as described in the above example command.

```
> MVPBT2(y,S)   # Generalized Egger test (MSSET2)
$T
          [,1]
[1,] 9.581602
```

```
$P
            [,1]
[1,] 0.008305801

$b0
[1]  0.6527201 -1.4483726
```

The *P*-value of MSSET2 was 0.008. However, it might be inaccurate because the uncertainty of the heterogeneity variance–covariance parameters was not considered.

The latter test (MSSET3) is more recommended in practice and is implementable by the following command:

```
> MVPBT3(y,S)    # Generalized Egger test (MSSET3)
$T.b
   [1]   3.728571970   0.143517438   0.648951676   2.418124253   0.659743986
1.280331731     1.215632849       0.479688613      0.395274721      2.343870993
1.917282832   3.404983696   0.667862531   3.146952026       ……

$T
         [,1]
[1,] 9.581602

$P
[1] 0.002998501
```

The *P*-value of MSSET3 was 0.004. Both of the generalized Egger tests indicated that the funnel plots in this example were asymmetric and the *P*-values were smaller than those of the univariate Egger tests. These *P*-values can be used in DTA meta-analyses, similar to the results of the ordinary Egger test in univariate meta-analysis.

An R file that includes the example code provided in this paper is available at [https://www.ism.ac.jp/~noma/MVPBT.r].

## 4. Concluding remarks

Evidence obtained from DTA meta-analyses has been widely used in public health, clinical practice, health technology assessments, and policy-making. However, publication biases such as those encountered in conventional univariate meta-analyses can lead to misleading evidence in these relevant applications. The new computational tools would be useful in detecting these potential biases and in circumventing misleading interpretations and conclusions. Also, more advanced methods and computational tools will be required in future methodological research to produce reproducible evidence.

## Funding

This study was supported by a Grant-in-Aid for Scientific Research from the Japan Society for the Promotion of Science (grant number: JP22H03554).

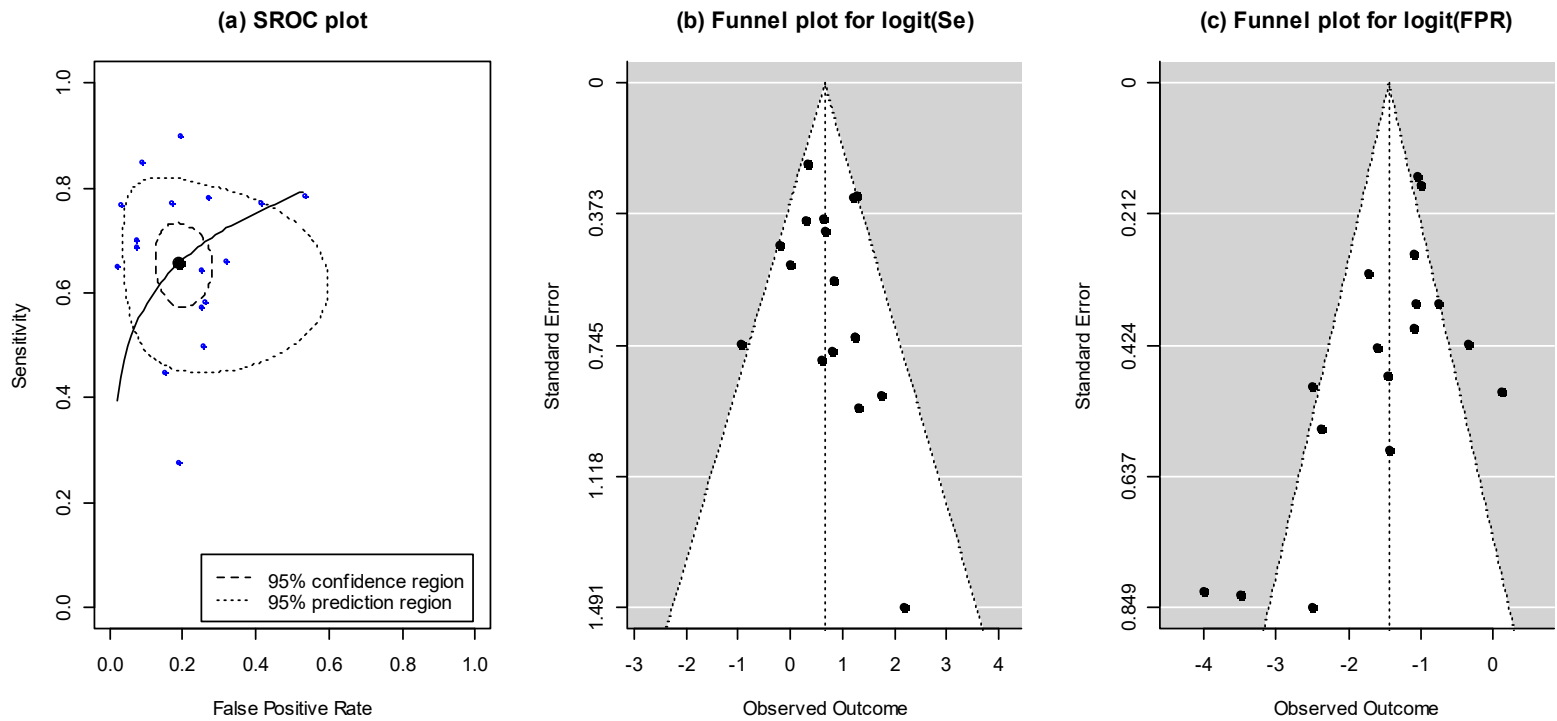

**Figure 1.** (a) Summary receiver operating characteristic (SROC) curve plot and (b, c) funnel plots for the marginal distributions of the logit-transformed sensitivity (logit(Se)) and the logit-transformed false-positive rate (logit(FPR)) of the DTA meta-analysis for lymphangiography.

**Table 1.** Non-technical descriptions of the publication bias tests involved in `MVPBT` package.

| Method / Function | Description |
| --- | --- |
| `MVPBT2` | `MVPBT2` is an efficient score test for the global null hypothesis of asymmetries of two funnel plots, incorporating the correlation information between the two outcomes. This test does not consider the uncertainty of the heterogeneity variance–covariance parameter estimates and therefore might be inaccurate under small $N$ (the number of studies) settings; Noma (2020) refers to this test as MSSET2. |
| `MVPBT3` | To address the uncertainty problem of the heterogeneity variance–covariance parameter estimates, Noma (2020) proposed a parametric bootstrap test for the efficient score test (MSSET2), where the bootstrap distribution of the score statistic is used instead of the ordinary $\chi^2$ distribution; Noma (2020) refers to this method as MSSET3. |

**Table 2.** Results of the DTA meta-analysis for lymphangiography.

|  | Summary estimate and 95% confidence interval |
|---|---|
| Sensitivity | 0.658 (0.590, 0.719) |
|   Between-studies SD † | 0.320 |
| False positive rate | 0.190 (0.135, 0.261) |
|   Between-studies SD † | 0.724 |
| AUC of the SROC curve | 0.755 |
|  | *P*-value for the univariate Egger test |
| Sensitivity | 0.551 |
| False positive rate | 0.029 |

† Between-studies SD estimates for logit-transformed outcomes.

**Table 3.** A summary of `edta` function.

| Argument | Description |
|---|---|
| TP | The number of true positives. |
| FP | The number of false positives. |
| FN | The number of false negatives. |
| TN | The number of true negatives. |

| Value | Description |
|---|---|
| y | Logit-transformed sensitivities and false positive rates. |
| S | Within-study variances and covariances. |
| Se | Sensitivities. |
| Fp | False positive rates. |

**Table 4.** A summary of `MVPBT2` and `MVPBT3` functions.

| Argument | Description |
| --- | --- |
| `y` | Summary outcome statistics (typically, logit-transformed sensitivities and false positive rates; easily computable by `edta` function). |
| `S` | Within-study variances and covariances of `y` (easily computable by `edta` function). |
| `B` (for `MVPBT3`) | Number of bootstrap resampling (default: 2000). |

| Value | Description |
| --- | --- |
| `T` | The efficient score statistic. |
| `P` | *P*-value of the publication bias test. |
| `b0` (for `MVPBT2`) | Constrained maximum likelihood estimates of the regression intercepts. |
| `T.b` (for `MVPBT3`) | Bootstrap samples of the efficient score statistic. |